\documentclass[aps,pre,groupedaddress,showkeys]{revtex4}
\usepackage{amsmath}
\usepackage{amssymb}
\usepackage{graphicx}

\begin{document}

\title{Modeling two-language competition dynamics}

\author{M.~Patriarca$^{(1,2)}$}\email{marcop@ifisc.uib-csic.es}
\author{X.~Castell\'o$^{(1)}$}\email{xavi@ifisc.uib-csic.es}
\affiliation{{\rm (1)} IFISC\thanks{\tt http://ifisc.uib-csic.es}, 
Instituto de F\'isica Interdisciplinar y Sistemas
  Complejos (CSIC-UIB),
  E-07122 Palma de Mallorca, Spain}\thanks{\tt http://ifisc.uib-csic.es}
\affiliation{{\rm (2)} National Institute of Chemical Physics and Biophysics,
R\"avala 10, 15042 Tallinn, Estonia}

\author{J.~R.~Uriarte}\email{jr.uriarte@ehu.es}
\affiliation{Universidad del Pais Vasco-Euskal Herriko Unibertsitatea,
Departamento de Fundamentos del An\'alisis Econ\'omico I,
Ekonomi Analisiaren Oinarriak I Saila, 48015 Bilbao, Spain}

\author{V.~M.~Egu\'iluz}\email{victor@ifisc.uib-csic.es}
\author{M.~San~Miguel}\email{maxi@ifisc.uib-csic.es}
\affiliation{IFISC, 
  Instituto de F\'isica Interdisciplinar y Sistemas
  Complejos (CSIC-UIB), 
  E-07122 Palma de Mallorca, Spain}\thanks{\tt http://ifisc.uib-csic.es}


\begin{abstract}
During the last decade, much attention has been paid to language
competition in the complex systems community, that is, how the
fractions of speakers of several competing languages evolve in time. In
this paper we review recent advances in this direction and focus on
three aspects. First we consider the shift from two-state models to
three state models that include the possibility of bilingual
individuals. The understanding of the role played by bilingualism is
essential in sociolinguistics. In particular, the question addressed
is whether bilingualism facilitates the coexistence of
languages. Second, we will analyze the effect of social interaction
networks and physical barriers. Finally, we will show how to analyze
the issue of bilingualism from a game theoretical perspective.
\end{abstract}


\maketitle


\section{Introduction}

The modeling of language dynamics in the general framework of complex
systems has been addressed from at least three different perspectives: 
language evolution (or how the structure of language
evolves)~\cite{Steels2011a}, language cognition (or the way in which
the human brain processes linguistic knowledge) 
\cite{Edelman2007a}, and language competition
(or the dynamics of language use in
multilingual communities)~\cite{Sole2010a,Stauffer2005a,Wichmann2008b}. 
The latter is the approach followed in this
paper in which, therefore, we focus on problems of social
interactions. 
Thus, there is a direct connection with social sciences and social dynamics,
since linguistic features represent cultural traits of a specific nature,
whose propagation and evolution can in principle be modeled through a
dynamics analogous to that of cultural spreading and opinion dynamics \cite{Castellano2009a,SanMiguel05a}.
Furthermore, language dynamics offers the possibility to
understand the mechanisms regulating the size evolution of linguistic
communities, an important knowledge in the design of appropriate linguistic policies.

This paper reviews recent work on language competition dynamics focusing on models describing interaction in social communities with two
languages or two linguistic features ``A'' and ``B''. 
In particular, we consider the
special role of bilingual speakers.
The emphasis is on two- or three-state models, in the physics jargon,
or models with two excluding or non-excluding options, to say it from the
perspective of economics and social norms.
Languages or linguistic features are considered here as fixed cultural traits in the society,
while the more general dynamics of evolving interacting languages is not
considered.

We discuss work done in several parallel directions and from different methodological points of view. A first approach reviewed below is the family of \emph{competition models}, referred to also as ``ecological models of language''.
They can be studied at a macroscopic or mesoscopic level,
introducing population densities for the respective language
communities or using microscopic agent-based models,
where the detailed interactions between agents are taken into account.
The formulation can be made at different levels of detail, possibly
introducing the effect of population dynamics, geography,
and intrinsic or external random fluctuations.
A different approach considered is that of game-theoretical models, which
allows the description of agents making choices about the language to be spoken at each encounter when
limited information is available to them. While in the ecological models each language is initially spoken by a fraction of the population, in the game-theoretical framework we consider the present situation in many societies in which one language is spoken by everyone, while a minority language is only spoken by a proportion of the population. The only relevant dynamics is then the one of the bilingual minority.

The outline of the paper is as follows: Section 2 describes language competition models. A first part is devoted to review the seminal model of Abrams and Strogatz \cite{Abrams2003a} and variations thereof, while a second part reviews several models proposed to take into account bilingual agents in this context. Section 3 discusses how to account for geographic effects in the models of language competition dynamics. Section 4 introduces a game-theory perspective into the problem. Section 5 contains some general conclusions and outlook.


\section{Language competition models}
\label{competition}

\subsection{The Abrams \& Strogatz model}
\label{secAS}

The Abrams \& Strogatz model (from now on, AS model)
\cite{Abrams2003a} is the seminal work which triggered a coherent
effort from a statistical physics and complex systems approach to the
problem of language 
competition~\footnote{See {\tt http://www.ifisc.uib-csic.es/research/complex/APPLET\_LANGDYN.html} for online applets of the AS model and the Bilinguals model considered in Sec.~\ref{secCastelloetal}.}.
It is a simple two-state model with two parameters (with a main focus
on prestige of the languages~\footnote{``Status'' is the term used in
  Ref.~\cite{Abrams2003a} to refer to prestige. However, in later
  publications authors have referred to it as ``prestige'', which
  appears to be more appropriate for sociolinguistic studies since in
  linguistics status usually refers to the degree of official recognition of a language.}), which the authors fit to real aggregated data of endangered languages such as Quechua (in competition with Spanish), Scottish Gaelic and Welsh (both in competition with English).
The original model is a population dynamics model. However in this review, we first introduce its microscopic version \cite{Stauffer2007a} and we consider later its population dynamics (mean-field) approximation.

The microscopic or individual based version of the AS model \cite{Stauffer2007a} is a two-state model proposed for the competition between two languages in which an agent {\it i} sits in a node within a network of $N$ individuals and has $k_{i}$ neighbors. Neighbors are here understood as agents sitting in nodes directly connected by a link. The agent can be in the following states: {\it A}, agent using language A (monolingual A); or {\it B}, agent using language B (monolingual B). At each iteration we choose one agent {\it i} at random and we compute the local densities for each of the states in the neighborhood of node {\it i}, $\sigma_{i, l}$ ({\it l}=$A$, $B$). The agent changes its state according to the following transition probabilities:
\begin{eqnarray}
p_{i, A \rightarrow B}=(1-s)(\sigma_{i, B})^a&,& \quad  \quad   p_{i, B \rightarrow A}=s(\sigma_{i, A})^a \hspace*{0.1cm}
\label{trans_prob_AS}
\end{eqnarray}
Equations~(\ref{trans_prob_AS}) give the probabilities for an agent
{\it i} to change from state {\it A} to {\it B}, or vice-versa. These
probabilities depend on the local densities ($\sigma_{i, A}$,
$\sigma_{i, B}$) and on two free parameters: the {\it prestige} of
language A, $0 \leq s \leq 1$ (the one of language B is $1-s$); and
the {\it volatility} parameter, $a \geq 0$. Prestige is modeled as a scalar which aggregates multiple factors. In this way, $s$ gives a measure of the different status between the two languages, that is, which is the language that gives an agent more possibilities in the social and personal spheres. Mathematically, it is a symmetry breaking parameter. The case of socially equivalent languages corresponds to $s=0.5$.
On the other hand, the volatility parameter gives shape to the functional form of the transition probabilities. The case $a=1$ is the neutral situation, where the transition probabilities depend linearly on the local densities. A high volatility regime exists for $a<1$, where the probability of changing language state is larger than the neutral case, and therefore agents change its state rather frequently. A low volatility regime  exists for $a>1$ with a probability of changing language state below the neutral case, and thus agents have a larger resistance to change its state. In this way, the volatility parameter gives a measure of the propensity (or resistance) of the agents to change their language use.

\subsubsection{Mean-field approximation}

In the limit of infinite population and fully connected society, that is, each agent can interact with any other agent in the population, the model can be described by differential equations for the population densities of agents \cite{Abrams2003a},
\begin{eqnarray} \label{AS}
  \dot{x}_A
  \!&=&\!
    {s_A} \, x_A^a \, x_B
  - {s_B} \, x_B^a \, x_A \ ,
  \nonumber \\
  \dot{x}_B
  \!&=&\!
  - {s_A} \, x_A^a \, x_B
  + {s_B} \, x_B^a \, x_A \ .
\end{eqnarray}
Here $x_i(t)$, $i=A,B$, represents the fraction of speakers of
language $i$ while $s_i$ represents the prestige.
Using the condition $x_A(t) + x_B(t) = 1$ and following the
normalization $s_A \!+\! s_B \!=\! 1$ ($s_A=s$), they can be
reformulated as a two-parameter single-variable problem for the variable $x_A(t)$:
\begin{equation}
\dot{x}_A = s_A \, x_A^a \, (1 - x_A) - s_B \, (1 - x_A)^a \, x_A~.
\end{equation}
The final state reached by the dynamics depends on the initial
populations $x_i(t_0)$ and the values of the parameters $s_i$ and $a$.
Note that from an ecological point of view the reaction term $R(x_A,x_B)={s_A} x_A^a x_B - {s_B} x_B^a x_A $
implies that languages A and B act to each other as preys and
predators at the same time~\cite{Murray2002a}, a situation peculiar in
ecology but realistic in the competition between cultural
traits~\cite{Abrams2003a}.

Abrams and Strogatz found an exponent $a=1.31$ when fitting to real
data from the competition between Quechua-Spanish, Scottish
Gaelic-English and Welsh-English~\cite{Abrams2003a}.
 Moreover, they also inferred the corresponding value of $s$ in each
 of the linguistic situations. The general analysis of the role of
 parameters $s$ and $a$ in the model is discussed below.

\begin{figure}[t]
 \includegraphics[width=0.4\textwidth]{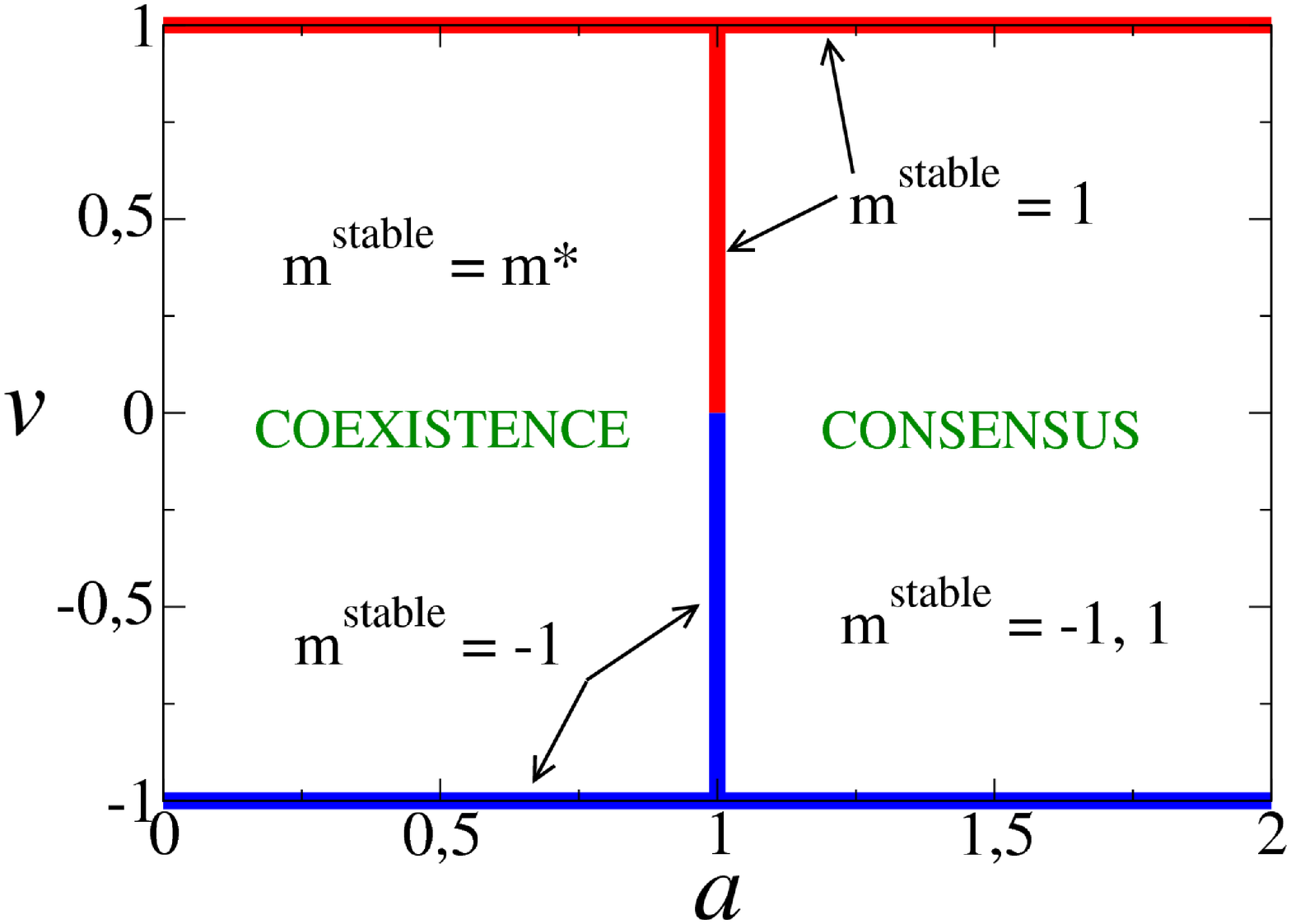}
 \includegraphics[width=0.4\textwidth]{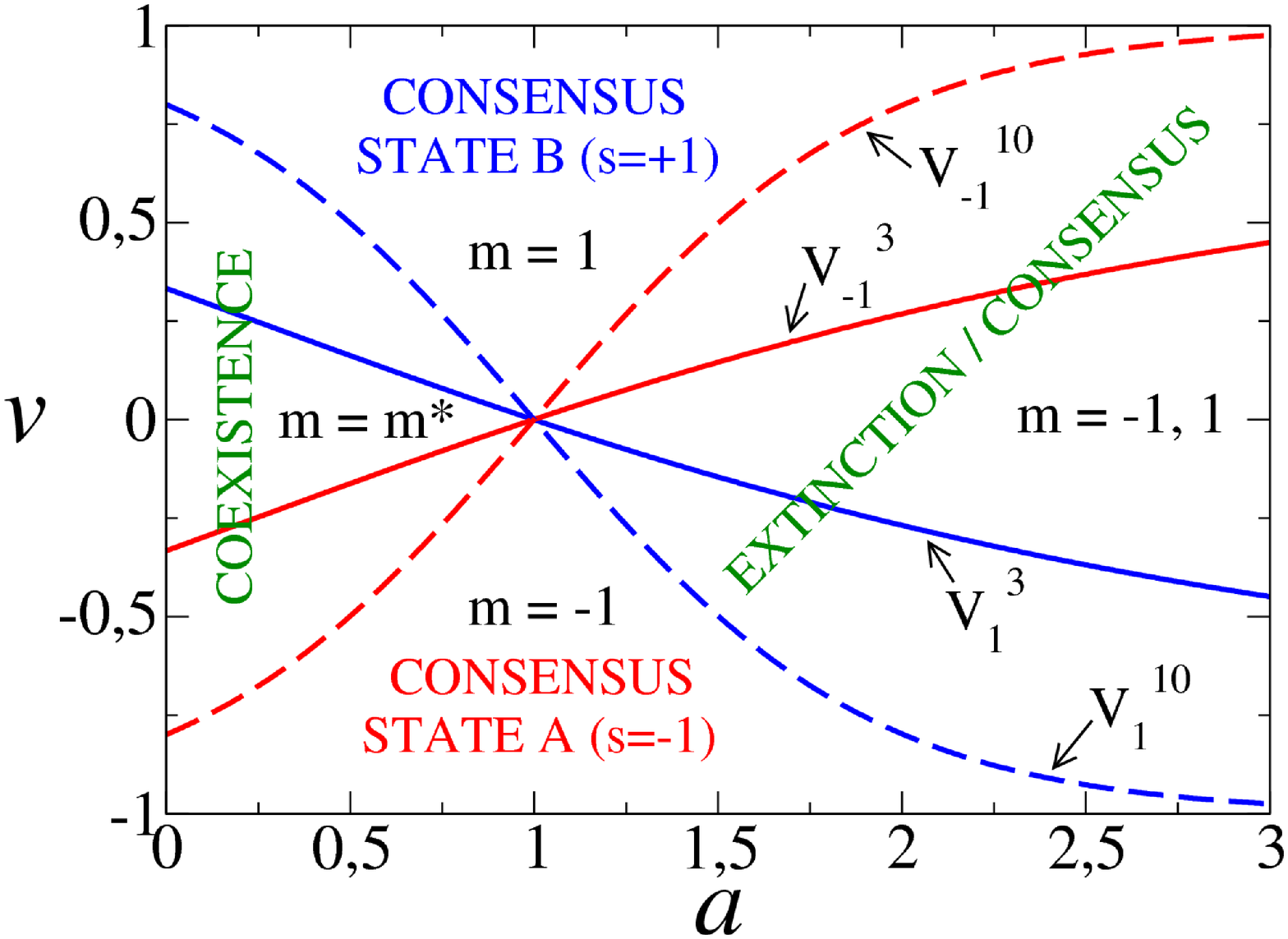}
 \caption { \label{stab-CG}
 Stability diagram for the AS model in a fully connected
 network (left) and a degree-regular random networks (right). Solid
 and dashed lines correspond to degree-regular random networks with
 degrees $\langle k \rangle=3$ and $\langle k \rangle=10$
 respectively. See the text for the definition of the regions. 
 From Ref.~\cite{Vazquez2010a}.}
\end{figure}

An alternative macroscopic description of the AS model can be obtained defining a \emph{magnetization} $m=x_A-x_B=2x_A -1$ and a bias parameter $v=1-2s$. The time evolution of the \emph{magnetization} is given by
\begin{equation}
\frac{dm(t)}{dt} = 2^{-(a+1)} (1-m^2) \left[ (1+v) (1+m)^{a-1} -(1-v)
  (1-m)^{a-1} \right].
\label{dmdt}
\end{equation}
Equation~(\ref{dmdt}) describes the evolution of a very large system
($N \gg 1$) at the macroscopic level, neglecting finite size
fluctuations. It has the three stationary solutions
\begin{eqnarray}
m_{-}=-x_B=-1, ~~~ m^*=\frac{(1-v)^{\frac{1}{a-1}}-
  (1+v)^{\frac{1}{a-1}}}{(1-v)^{\frac{1}{a-1}}+(1+v)^{\frac{1}{a-1}}}~~~
\mbox{and}~~~ m_{+}=x_A=1.
\label{solutions_AS}
\end{eqnarray}
Solutions $m_{\pm}$ correspond to the dominance of one of the
languages, so that a social consensus on which language to be used has
been reached. Solution $m^*$ corresponds to a situation of coexistence
of the two languages. When $a<1$ (large volatility), both solutions
$m_{\pm}$ are unstable, and $m^*$ is stable, whereas for $a>1$ (small
volatility) the opposite happens.
In the line $a=1$, $m_{+}$ is unstable (stable) for $v<0$ ($v>0$), and vice-versa
for $m_{-}$. Therefore, a structural transition is found at the critical value $a=1$.
In Fig.~\ref{stab-CG}, we show the regions of stability and instability of
the stationary solutions on the $(a,v)$ plane obtained from the above
analysis. We observe a region of coexistence ($m^*$ stable) and one of
bistable dominance of any of the languages ($m_{+}$ and $m_{-}$ stable) for any value of the prestige parameter.

The AS model has been studied within the context of viability theory and resilience \cite{Aubin_1991,Chapel2010a}. In this framework, it is assumed that the prestige $s$ of the model can be changed in real time by action policies in order to maintain the coexistence of two competing languages, that is, keeping language coexistence viable. In Ref.~\cite{Chapel2010a} such policies are obtained, studying the effect of varying the volatility parameter. In general, large values of $a$ (small volatility) reduce the set of viable situations for language coexistence.

\subsubsection{Random networks}

In order to account for local effects in which each agent only
interacts with a small fraction of the population one needs to go
beyond the mean field approximation. A first step in this direction is
considering the AS model in random networks. For these networks a pair
approximation can be used in which the state of an agent only depends
on the state of its first neighbors. This approximation works well in
networks with no correlations to second neighbors. For a
degree-regular random network, that is, a network in which each 
node is randomly connected to a fixed number of $\langle k \rangle$ neighbors,
the phase diagram obtained in Ref.~\cite{Vazquez2010a} is shown in
Fig.~\ref{stab-CG}. Compared to the fully connected case, the region
of coexistence is found to shrink for $v \neq 0$, as there appear two
regions where only the solution corresponding to dominance of the most
prestigious language is stable (labeled as CONSENSUS STATE A
[B]). 
These regions become larger, the smaller the average number of
neighbors $\langle k \rangle$, because the local effects become more
important (see transition lines for $\langle k \rangle=3$ and $\langle
k \rangle=10$ in Fig.~\ref{stab-CG}), reducing further the set of parameters
for which coexistence is possible. 

These regions become larger as local effects in the network
become more important as the number of neighbors $\langle k \rangle$
diminish (see transition lines for $\langle k \rangle=3$ and $\langle
k \rangle=10$ in Fig.~\ref{stab-CG}), reducing further the set of parameters
for which coexistence is possible. 
The fourth region
(EXTINCTION/CONSENSUS) corresponds to situations where dominance in
any of the two languages is stable.

\subsubsection{Two-dimensional lattices}

When two-dimensional lattices are considered, correlations to second
neighbors become important. Thus a pair approximation is not enough to
describe the system macroscopically. Instead, a field approximation is
needed~\cite{Vazquez2008c}, which describes well the macroscopic
evolution of the system and is able to characterize the different
dynamics of the growth of linguistic domains depending on the
volatility parameter, $a$. The stability diagram obtained for a square
lattice is qualitatively similar to the one found in random networks
(Fig.~\ref{stab-CG}), but the region for coexistence is found to be
much narrower than the ones observed in complex networks with low
degree (small number of neighbors).

In the field approximation \cite{Vazquez2010a}, we define $\phi_{\bf r}(t)$ as the average \emph{magnetization} field at site ${\bf r}$ at time $t$, which is a continuous representation of the \emph{magnetization} at that site ($-1<\phi<1$). In this description, an alternative and more visual way of studying stability is by writing the time evolution of the \emph{magnetization} field $\phi_{\bf r}$ in the form of a time-dependent Ginzburg-Landau equation
\begin{equation}
\frac{\partial \phi_{\bf r}(t)}{\partial t} = D(\phi_{\bf r}) \Delta
\phi_{\bf r} - \frac{\partial V_{a,v}(\phi_{\bf r})}{\partial
  \phi_{\bf r}}~,
\end{equation}
with potential $V_{a,v}(\phi_{\bf r})$ and diffusion coefficient $D(\phi_{\bf r})$.

The study of the potential allows us to obtain the different regimes of domain growth of the model. An especially interesting case is that of $v=0$ (socially equivalent languages). For $a<1$ there is no domain growth, and a large system does not order remaining in a dynamically changing state of language coexistence. The value $a=1$ appears to be critical, as for this value the potential becomes $V_{a,v}(\phi_{\bf r})=0$, and then linguistic domains grow by interfacial noise (Voter model dynamics) leading to a final ordered absorbing state of dominance of one of the languages. For $a>1$ instead, the mechanism of domain growth changes: it is driven by surface tension, and the system also orders leading to a state of dominance of one of the languages.
Notice that the order-disorder non-equilibrium transition at $a_c=1$ is of first order.

\subsection{Introducing bilingual speakers}
\label{sec:Bil}

\subsubsection{Minett \& Wang model}
\label{secMW}

 Minett \& Wang proposed a natural extension of the AS model in which bilingual agents are introduced in the dynamics. They made a schematic proposal of how to include such agents in Ref.~\cite{Wang2005a}. After a first proposal in a working paper in 2005, they published a model which considers both vertical and horizontal transmission \cite{Minett2008a}. This model focuses on language competence rather than use, and has seven free parameters, including prestige, volatility, four different peak transition rates, and a mortality rate \cite{Minett2008a}. However, they only present results regarding the case of neutral volatility ($a=1$).

In this context, the first relevant results concern a dynamical systems approach of the model, fixing most of the parameters to given values and analyzing mainly the role of the prestige parameter. This includes a stability analysis of the fixed points and the study of the basins of attraction in phase portraits. They propose a simple language policy which consists in changing the prestige of a language once the total density of speakers falls below a given value (intervention threshold) for which the language is considered to be in danger~\footnote{Notice that differently to this policy, when using viability theory \cite{Castello2011a,Chapel2010a} we suppose that the prestige can take any value although the action on the prestige is not immediate: the time variation of the prestige is bounded.}. In this way, they show a possible coexisting scenario for both languages.

The second important result concerns the study of an agent based model, which takes into account a discrete society. They analyze fully connected networks and the so-called local world networks \cite{Li_Chen_2003} (where agents link by preferential attachment only to a subset of the total number of nodes in the network) for which they analyze the frequency of convergence to each of the equilibria depending on the intervention threshold.
They obtain similar results for the dynamics of both networks when no language policy favoring coexistence is applied; but they found that once this policy takes place, maintenance is more difficult in local world networks.

\subsubsection{Castell\'o et al. model}
\label{secCastelloetal}

The work by Castell\'o et al. aims to study the dynamics of language competition taking into account complex topologies of social networks, finite size effects and different mechanisms of growth of linguistic domains. Their Bilinguals model is an extension of the AS model inspired in the original proposal of Minett and Wang. In this model agents can also be in a third bilingual state, $AB$, where agents use both languages, A and B. There are three local densities to compute for each node {\it i}\,: $\sigma_{i, l}$ (${\it l}=A, B, AB$). An agent  $i$ changes its state according to the following transition probabilities:
\begin{eqnarray}
p_{i, A \rightarrow AB}=(1-s)(\sigma_{i, B})^a&,& \quad  \quad   p_{i, B \rightarrow AB}=s(\sigma_{i, A})^a, \hspace*{0.1cm}
\label{trans_prob_MW_1}
\\
p_{i, AB \rightarrow B}=(1-s)(1-\sigma_{i, A})^a&,&  \quad  \quad  p_{i, AB \rightarrow A}=s(1-\sigma_{i, B})^a,
\label{trans_prob_MW_2}
\end{eqnarray}
which depend on the same two parameters of the AS model: prestige ($s$) and volatility ($a$).
Equations~(\ref{trans_prob_MW_1}) give the probabilities for changing from a monolingual state, $A$ or $B$, to the bilingual state $AB$, while equations~(\ref{trans_prob_MW_2}) give the probabilities for an agent to move from the $AB$-state towards the $A$ or $B$ states. Notice that the latter depends on the local density of agents using the language to be adopted, including bilinguals ($1-\sigma_{i, l}=\sigma_{i, j} + \sigma_{i, AB}$, ${\it l, j}=A,B$; $l\neq j$). It is important to stress that a change from state $A$ to state $B$ or vice-versa always implies an intermediate step through the $AB$-state~\footnote{Notice that in the analysis of the AS model and the Bilinguals model, the {\it use} of a language rather than the {\it competence} is considered. In this way, learning processes are out of reach of the present models. Effectively, the situation is such as if all agents were competent in both languages.} .

In the mean field limit, the model is described by the following differential equations for the total population densities of agents $x_{A}, x_{B}$ ($x_{AB}=1-x_{A}-x_{B}$),
\begin{eqnarray}
{\rm d}x_{A}/{\rm d}t = s(1-x_{A}-x_{B})(1-x_{B})^{a}-(1-s)x_{A}(x_{B})^a \hspace*{0.1cm},
\label{BM_mean_field_EQ_1}
\\
{\rm d}x_{B}/{\rm d}t = (1-s)(1-x_{A}-x_{B})(1-x_{A})^{a} - s(x_{A})^ax_{B}.
\label{BM_mean_field_EQ_2}
\end{eqnarray}
Equations~(\ref{BM_mean_field_EQ_1})-(\ref{BM_mean_field_EQ_2}) have three fixed points: $(x_{A},x_{B},x_{AB})=(1,0,0),(0,1,0)$, which correspond to consensus in the state A or B respectively; and $(x_{A}^{*},x_{B}^{*},x_{AB}^{*})$, with $x_{l}^{*}\neq 0$ (${\it l}=A,B,AB$).
There are no closed expressions for $x_{l}^{*}$ (${\it l}=A,B,AB$) and numerical analyses are needed. The dynamics of the AS model and the Bilinguals model in the whole parameter space has been analyzed in detail in Ref.~\cite{Vazquez2010a}, where macroscopic descriptions are obtained for fully connected networks, random networks and two-dimensional lattices; and order-disorder transitions are found and analyzed in detail (asymptotic states).

When introducing bilingual agents, the order-disorder transition described in Sec.~\ref{secAS} (Fig.~\ref{stab-CG}) is qualitatively the same, but the whole stability diagram shifts towards smaller values of the parameter $a$ \cite{Vazquez2010a}. In fully connected networks, the critical value shifts from $a_c=1$ to $a_c=0.63$. In random networks and square lattices the whole stability diagram shifts to smaller values of the parameter $a$, with $a_c(v=0)=0.3$ and $a_c(v=0)=0.16$ respectively.
Therefore, bilingual agents are found to generally reduce the scenario of language coexistence in the networks studied.

{\bf Socially equivalent languages and neutral volatility.} The role
played by bilingual agents in the dynamics of language competition
becomes more evident when considering the particular case of socially
equivalent languages ($s=0.5$) and neutral volatility ($a=1$), in
which the AS model reduces to the Voter model
\cite{Holley1975,Ligget_1999,Vazquez2008a}, and the Bilinguals model to the AB model \cite{Castello2006a,Castello2007a}.

\begin{figure}[tb]
\vspace{0.2cm}
\includegraphics[width=0.4\textwidth]{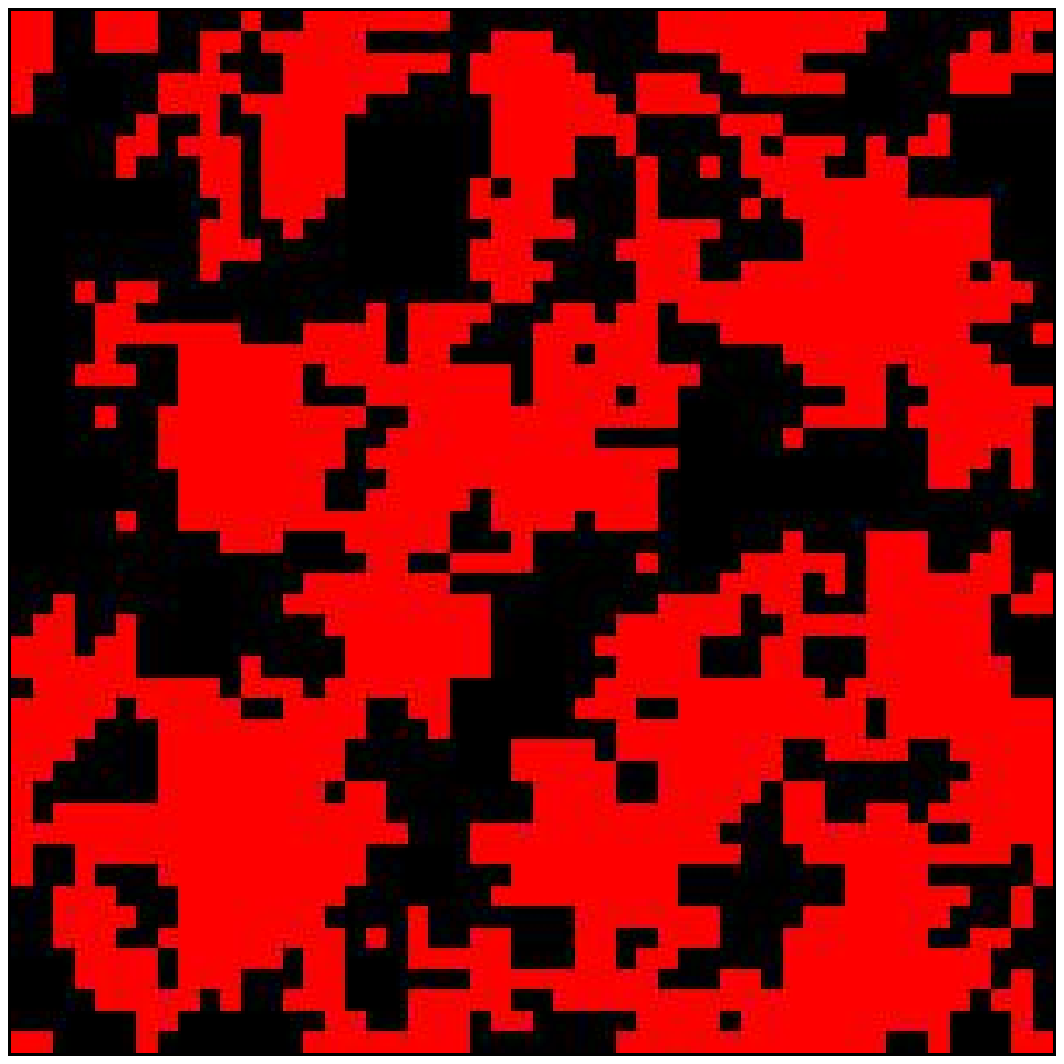},
\includegraphics[width=0.4\textwidth]{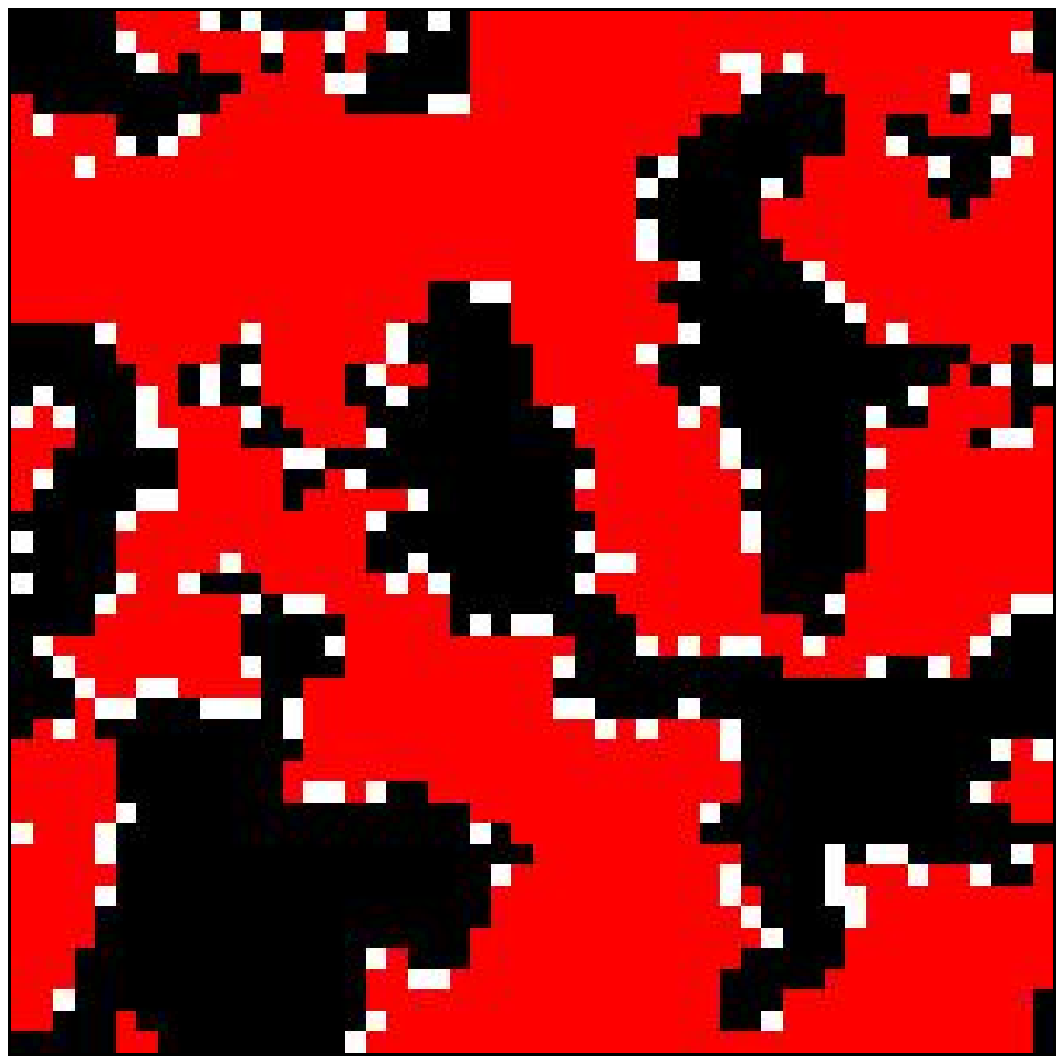}
\caption {\label{fig:ssnap}
Snapshots of the Voter model (left) and the Bilinguals model
  (right). Speakers of language A (red), B (black) and bilinguals
  (white) after 80 time steps starting with a random initial
  condition.}
\end{figure}

A first relevant result concerns the different interface dynamics for the growth of linguistic domains observed on two-dimensional lattices. The addition of the third intermediate state (bilingual agents) results in a change of the interfacial noise dynamics characteristic of the Voter model to a curvature driven dynamics, characteristic of spin flip Kinetic Ising  dynamics \cite{Gunton_1983}, changing the growth of monolingual spatial domains \cite{Castello2006a,Castello2011a} (see Fig.~\ref{fig:ssnap}). The time evolution of the characteristic length of a domain $\xi(t)$ changes from $\xi \sim \ln(t)$ to $\xi \sim t^{\alpha}$, with $\alpha\simeq 0.5$.
In addition, bilingual domains are never formed. Bilingual agents place themselves at the boundaries between monolingual domains.
These results imply that the AB model behaves as a local majority model with two states ($A$ and $B$), with bilingual agents at the interfaces.
As we discuss in the next paragraphs, this change in the interface dynamics turns to be crucial in the different behavior of the Bilinguals model observed in different networks when compared to the AS model.

The second result is related to the role of social networks of increasing complexity.
In the first place, small world networks \cite{Strogatz1998} are considered, which take into account the existence of long range interactions throughout the network.
In comparison to the Voter model, where the dynamics reaches a
metastable state~\footnote{Notice that the critical dimension for the
  Voter model is $d=2$. Therefore, in complex networks the system falls
  in metastable dynamical states which only reach an absorbing state
  (language dominance) due to finite size fluctuations.}, bilingual
agents restore the processes of domain growth and they speed-up the
decay to the absorbing state of language dominance by finite size
fluctuations \cite{Castello2006a}. The characteristic time, $\tau$ to
reach an absorbing state scales with the rewiring
parameter~\footnote{A small world network occurs for intermediate
  values of $p$ between $p=0$ (regular network) and $p=1$ (random network).} as $\tau \sim p^{-0.76}$.

\begin{figure}[tb]
\vspace{0.2cm}
\includegraphics[width=0.48\textwidth]{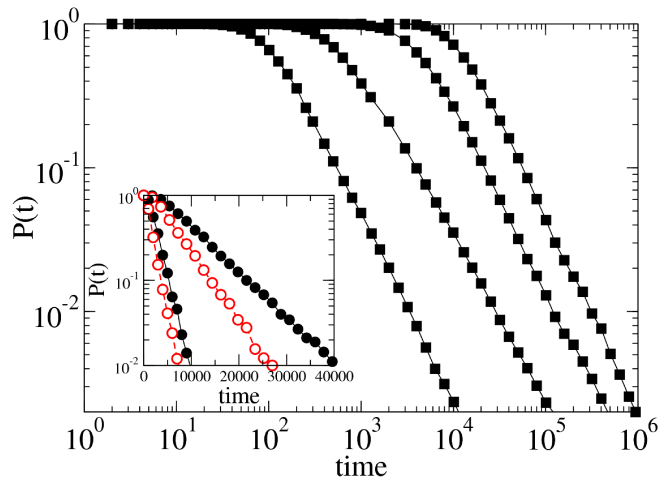}
\caption {\label{fig:aliveruns}
Fraction of alive runs in time for networks 
with communities (solid symbols) 
for the AB model (double logarithmic plot).
The system sizes are $N= 100$,
$400$, $2500$, $10000$ from left to right, with averages taken over
different realizations of the network ($400$-$5000$ depending
on system size), with $10$ runs in each. 
Inset: same plot for networks 
with communities (solid symbols) and 
randomized networks (empty symbols) for the Voter model
(semilogarithmic plot).
The system sizes are $N= 2500$, $10000$. 
Averages are taken over $100$ different
realizations of the networks, 
with $10$ runs in each. From Ref.~\cite{Castello2007a}. }
\end{figure}

Secondly, networks with community structure are analyzed, following the algorithm by Toivonen et al.~\cite{Toivonen_2006}. These networks mimic most of the features of real social networks: the presence of hubs, high clustering, assortativity, and mesoscale structure. Communities do not affect substantially the Voter model dynamics, where the system reaches again metastable states. Instead, the presence of communities dramatically affects the AB model \cite{Castello2007a}. On the one hand, linguistic domains correlate with the community structure, with bilingual agents connecting agents belonging to different communities, leading to trapped metastable states. The analysis of the lifetime distributions to reach an absorbing state shows that there is no characteristic time for the dynamics (Fig.~\ref{fig:aliveruns}): trapped metastable states are found at arbitrary long times, which lead to scenarios of long time language segregation. Notice that the change in the interface dynamics found in two-dimensional lattices mentioned above when introducing bilingual agents, from interfacial noise to curvature reduction, is at the basis for the change of behavior observed in complex networks .

The AB model has also been compared to the Naming Game restricted to two conventions (\emph{2c-Naming Game}) \cite{Baronchelli2007a,Castello2009a}.
The general Naming Game \cite{Steels1996} describes a population of agents playing pairwise interactions in order to {\em negotiate} conventions, i.e., associations between forms and meanings, and elucidates the mechanisms leading to the emergence of a global consensus among them.
In the case of two conventions, the Naming Game can be also interpreted as a language competition model with two non-excluding options.
The AB model and the 2c-Naming Game are found to be equivalent in the mean field approximation.
However, the main result concerns the fact that, when these are
extended incorporating a parameter $\beta$ which describes the inertia
of the agents to abandon an acquired language, 
they show an important difference with respect to the existence of an
order-disorder transition.
While the 2c-Naming Game features
an order-disorder transition between consensus and stationary
coexistence of the three phases present in the system \cite{Baronchelli2007a}, in the
AB model such a transition does not exist \cite{Castello2009a}.

\subsubsection{Mira et al. model}
\label{secMiraetal}

The works of Mira et al.~\cite{Mira2005a,Mira2011a}
introduce the category of bilinguals,
in the particular but relevant case in
which two languages are relatively similar to each other and
partially mutually intelligible.
In such a situation any speaker of e.g. the speaking community B can
learn the language of the other speaking community A with a small
effort, thus becoming a member of the bilingual community AB.
In practice, this is described in the model through a transition
probability, to switch from monolingual B to a bilingual community AB,
which is relatively large compared to that of the opposite transition,
as discussed below.
This changes the dynamics in such a way that stable coexistence is
possible.

The model may actually be given more general interpretations,
since there may be situations in which the transition from a monolingual
community B to the bilingual community AB is very probable or easiest to
carry out for reasons other than linguistic similarity,
e.g. tight economical needs which requires both
languages or social policies which favor or push B speakers to learn
language A.

The model describes the time evolution of
the fractions $x_A$ and  $x_B$
of speakers in the monolingual communities A and B and
the analogous fraction $x_{AB}$ for the bilingual community AB.
In the approximation of a constant total population,
$x_{A} + x_{B} + x_{AB} = 1$,
one can eliminate the variable $x_{AB}$, obtaining
\begin{eqnarray} \label{MP}
  \dot{x}_{A}
  \!&=&\!
    (1 - x_{A}) (1 - k) s_{A} (1 - x_{B})^a
  - x_{A} (1 - s_{A}) (1 - x_{A})^a \, ,
  \nonumber \\
  \dot{x}_{B}
  \!&=&\!
    (1 - x_{B}) (1 - k) (1 - s_{A}) (1 - x_{A})^a
  - x_{A} s_{A} (1 - x_{B})^a \, ,
\end{eqnarray}
which reduce to Eqs.~(\ref{AS}) of the AS model for $k=0$.
Here the normalization for the language status $s_{A} + s_{B}
= 1$ has been assumed, so that the actual parameters of the model are
$a$, $k$ and $s_{A}$.
The closer the parameter $k$ is to one, the more probable is the
transition from monolingual B to bilingual AB.
Different types of stable equilibrium states have been shown to
exist in the phase-plane of the model~\cite{Mira2005a,Mira2011a},
including some with non-zero bilingual
community size $x_{AB} > 0$ even for $s_{A} \ne 1/2$,
i.e. when one of the languages is favored by a higher status.
In order for this condition to happen 
a sufficiently large value of $k$ is needed~\cite{Mira2011a}.

\subsubsection{Lotka-Volterrra like models}
\label{secPR}

In Ref.~\cite{Pinasco2006a} a Lotka-Volterra-like
dynamical model is introduced, which provides a
counter-example to the conclusions of the AS model
(which predicts the extinction of one of the two competing languages
in a broad range of values of the prestige $s$ and volatility $a$ parameters).
A main ingredient of the model is the introduction of population dynamics with
different Malthus rates $\alpha_{A}$ and
$\alpha_{B}$ for the two populations A and B.
The dynamical equations read
\begin{eqnarray}
 \label{PR}
  \dot{x}_{A}
  \!&=&\!
    c \, x_{A} x_{B}
    + \alpha_{A} x_{A} \left(1 -
    \frac{x_{A}}{K_{A}} \right) \, ,
  \nonumber \\
  \dot{x}_{B}
  \!&=&\!
    - c \, x_{A} x_{B}
    + \alpha_{B} x_{B} \left(1 -
    \frac{x_{B}}{K_{B}} \right) \, ,
\end{eqnarray}
where $c$ measure the switch rate from language B to language A,
while the carrying capacities $K_{A}$ and $K_{B}$
represent the maximum population sizes allowed in case of isolated
populations ($c = 0$).
The language with lower status --- in the present case language B
since it is assumed that $c > 0$ --- is shown to be able to survive,
as long as its corresponding reproduction rate is
higher than its language disappearance rate, i.e. if
$c K_{B} < \alpha_{B}$.
This system presents a stable equilibrium solution in which both
communities survive ($x_{A}, x_{B} > 0$).

As noticed in Ref.~\cite{Kandler2008a}, the equilibrium value of the
population size $x_{A}$ is larger than the maximum allowed value
$K_A$ for population A.
In order to overcome this problem, Kandler et
al.~\cite{Kandler2008a} have proposed some different models.
In particular, they introduced a model with a common carrying capacities
$K$ for the two language communities, in the sense that in principle
both the communities could reach the same maximum population size $K$.
To take into account the fact that in practice the resources actually
available to a community are limited by those already used by the other
community, their model (in its zero-dimensional version) is defined
by the following equations,
\begin{eqnarray} \label{KS1}
  \dot{x}_{A}
  \!&=&\!
    c \, x_{A} x_{B}
    + \alpha_{A} x_{A}
      \left(
      1 - \frac{x_{A}}{K - x_{B}}
      \right) \, ,
  \nonumber \\
  \dot{x}_{B}
  \!&=&\!
    - c \, x_{A} x_{B}
    + \alpha_{B} x_{B}
      \left(
      1 - \frac{x_{B}}{K -  x_{A}}
      \right) \, .
\end{eqnarray}
Notice that while the transition between A and B communities is
regulated by the same Lotka-Volterra-type rate of the model described
in Eq.~(\ref{PR}), the Verhulst terms in the population
dynamics part of the equations now contain the effective carrying
capacities $K_{A}(t) = K - x_{B}(t)$ and the
analogous one for population B, which prevent the population sizes from
overcoming the carrying capacity $K$.
However, with this change the coexistence equilibrium state is lost
again and only one community can survive asymptotically.
The addition of heterogeneity to the model can change things substantially, as
discussed below in Sec.~\ref{secGeography}.

The model was then generalized to include the bilingual community
size~\cite{Kandler2009a,Kandler2008a}.
The transition dynamics is similar to that of the
Minett and Wang model but effective carrying capacities of the population dynamics part are
designed in a way similar to those
used above in Eqs.~(\ref{KS1}).
It was still found that the presence of bilingualism does not allow the
coexistence of two languages asymptotically and that bilinguals
represent the interface between the two monolingual communities;
however, the presence of bilinguals can in some cases prolong
significantly the extinction time.


\section{Models with geography}
\label{secGeography}

The importance of physical geography, e.g. the presence of
water boundaries and mountains, for the evolution and
dispersal of biological species is well known~\cite{Lomolino2006a}.
In cultural diffusion, both physical and political boundaries have
to be taken into account.
Furthermore, other factors of dynamical or economical nature, related
to e.g. to the features of the landscape, can modify the otherwise
homogeneous cultural spreading process.

The geographical models considered
below~\cite{Kandler2008a,Patriarca2009a,Patriarca2004c} are
extensions of the 0-dimensional AS model to spatial domains.
In all cases some geographical inhomogeneities are taken into account,
related to the underlying political, economical, or physical geography.

\subsection{Inhomogeneous cultural diffusion}
\label{sec:PL}

Inhomogeneous cultural diffusion can be modeled by adopting space
(and time) dependent transition rates for the switching between
different languages.

In the model introduced in Ref.~\cite{Patriarca2004c}
the speakers of two communities with different languages A and B
can diffuse \emph{freely} across a two-dimensional domain $Z$,
divided symmetrically into two regions $\alpha$ and $\beta$.
The border between the regions, which could represent e.g. political or
geographical factors, is assumed to influence the communication
between speakers in such a way that language A is
more influential in zone $\alpha$ and language B in zone $\beta$.
The reaction dynamics is similar to that of the AS model, with the
crucial difference that (1) there is free homogeneous diffusion
through the domain $Z = \alpha + \beta$ and (2) each speaker is only
affected by the other speakers who are in the same region where the speaker is.
The AS dynamics is modified as follows,
\begin{eqnarray}
  \frac{\partial f_{A}({\bf r},t)}{ \partial t }
   &=& \Delta f_{A}({\bf r},t) + R({\bf r},t) \ ,
  \nonumber \\
  \frac{\partial f_{B}({\bf r},t)}{ \partial t }
   &=& D \Delta f_{B}({\bf r},t) - R({\bf r},t) \ ,
  \nonumber \\
   R({\bf r},t)
   &=&
   \left[
     s_{A} \, \xi_{A}^{\,a}({\bf r},t) f_{B}({\bf r},t)
     - s_{B} \, \xi_{B}^{\,a}({\bf r},t) f_{A}({\bf r},t)
   \right]
  \ ,
  \label{f}
\end{eqnarray}
where $\mathbf{r} = (x,y)$, $\Delta$ is the Laplacian in two
dimensions, $D$ the diffusion coefficient,
$f_{A}$ and  $f_{B}$ the population
density of the speaking community A and B, respectively, and $s_i$ ($i
= {A}, {B}; s_{A} + s_{B} = 1$) maintain the same meaning
of the language status.
The terms $\xi_i({\bf r},t)$ in the reaction rate $R({\bf r},t)$ are the
fractions of speakers of language $i$ at time $t$ in the same region
($\alpha$ or $\beta$) where a speaker with position ${\bf r}$ is,
\begin{eqnarray}
  \xi_i({\bf r},t)
  &=& { N_i^{(\alpha)}({\bf r},t)}
   /
  [ N_{A}^{(\alpha)}({\bf r},t) + N_{B}^{(\alpha)}({\bf r},t) ]
  ~~~~\mathrm{if}~\mathbf{r} \in \alpha \, ,
  \\
  &=& { N_i^{(\beta)}({\bf r},t)}
  /
  [ N_{A}^{(\beta)}({\bf r},t) + N_{B}^{(\beta)}({\bf r},t) ]
  ~~~~\mathrm{if}~\mathbf{r} \in \beta \, ,
  \\
  N_i^{(\alpha)}({\bf r},t)
  &=& \int_\alpha dx' dy' f_i({\bf r}',t) \, ,
  ~~
  N_i^{(\beta)}({\bf r},t)
  = \int_\beta dx' dy' f_i({\bf r}',t) \, ,
  ~~i = {A}, {B} \, .
  \label{N_mu}
\end{eqnarray}
The model exhibits stable equilibrium with both languages surviving in
the two different regions $\alpha$ and $\beta$, also for very different
language status, if the two populations are initially separated on the
opposite sides of the boundary.

Another example of inhomogeneous language spreading modeling is
provided by the model introduced in Ref.~\cite{Kandler2008a}, where
the transition rate $c$ in Eqs.~(\ref{KS1}) is allowed to vary in
space, i.e. $c \to c({\bf r})$, and diffusion is taken into account.
This analogously models a situation in which the two languages have a
higher status in different spatial domains.
The equations read
\begin{eqnarray}
  \frac{\partial f_{A}({\bf r},t)}{ \partial t }
   &=&
    D_{A} \Delta f_{A}({\bf r},t)
    + c({\bf r}) f_{A}({\bf r},t) f_{B}({\bf r},t)
    +  \alpha_{A} f_{A}({\bf r},t)
       \!\!\left[
       1 - \frac{f_{A}({\bf r},t)}{K \!-\! f_{B}({\bf r},t)}
    \right]\!,~~~~
  \nonumber \\
  \frac{\partial f_{B}({\bf r},t)}{ \partial t }
   &=&
    D_{B} \Delta f_{B}({\bf r},t)
    - c({\bf r}) f_{A}({\bf r},t) f_{B}({\bf r},t)
    +  \alpha_{B} f_{B}({\bf r},t)
    \!\!\left[
    1 - \frac{f_{B}({\bf r},t)}{K \!-\! f_{A}({\bf r},t)}
    \right]\!\!.~~~~
  \label{f2}
\end{eqnarray}
The model predicts survival of both languages in the two different
regions for suitable forms of the function $c({\bf r})$.

It is noteworthy that Kandler et al. applied their geographical models
in real situations,
to test different social strategies planned for defending the survival
of Britain's Celtic
languages~\cite{Kandler2009a,Kandler2008a,Kandler2010a}.
For further details and data source references see
Ref.~\cite{Kandler2010a}.
It is also worth pointing out that the geographical character of these
models may be given a wider interpretation in terms of social space.
For instance, the $x$ coordinate may represent age, different income
classes, or different social environments 
(e.g. workplace versus administration, workplace versus home, etc.).
Thus, while in all these cases there is actual competition between languages
everywhere, in practice each language eventually may turn up to be successful
in and therefore characterize a different specific social domain.
The diversity of different languages or language features
or their ability 
to find a suitable niche where to be successful is crucial to the
survival of any minority language.

\subsection{Inhomogeneous human dispersal}
\label{sec:PH}

The model introduced in Ref.~\cite{Patriarca2009a} can be considered to
be complementary to those discussed in the previous section, 
in the sense that it assumes a
homogeneous switching rate in parallel with an inhomogeneous diffusion.
It can be noticed that while inhomogeneous diffusion is typically
caused by factors related to the physical geographic features of the
underlying landscape, it can also be due e.g. to political boundaries
or natural borders between regions with very different economical
features.
It turns out that even such non-cultural features can strongly
influence the evolution and final distribution of cultural traits.

In the continuous limit the model assumes a dynamics of a
reaction-diffusion type, described by
\begin{eqnarray}
   \frac{\partial f_1}{ \partial t }
   &=&
   R(f_1,f_2)
   - \nabla \cdot \left( \, \mathbf{F} f_1 \right)
   +  \nabla  \cdot ( D \nabla f_1 )
   + \alpha f_1 \left( 1 - \frac{f_1+f_2}{K} \right) \ ,
   \label{f2a}
   \\
   \frac{\partial f_2}{ \partial t }
   &=&
   - R(f_1,f_2)
   - \nabla \cdot \left( \, \mathbf{F} f_2 \right)
   + \nabla  \cdot ( D \nabla f_2 )
   + \alpha f_2 \left( 1 - \frac{f_1+f_2}{K} \right) \ ,
   \label{f2b}
   \\
   R(f_1,f_2)
   &=&
   k \left( s_1 f_1^{\, a}  f_2 - s_2 f_2^{\, a}  f_1 \right) \, .
\label{f2c}
\end{eqnarray}
Here the first term on the right hand side of Eqs.~(\ref{f2a}) and
(\ref{f2b}) is the reaction term $R$, which can be recognized from
Eq.~(\ref{f2c}) to be formally identical to that of the AS
model, if the population fractions are replaced by the population
densities.
Inhomogeneous diffusion can be due to the advection term containing
the external ``force field'' $\mathbf{F}(x,y) = (F_x(x,y), F_y(x,y))$
as well as to an inhomogeneous diffusion coefficient $D=D(x,y)$.
The model also contains logistic terms with Malthus rate $\alpha$ and
carrying capacity $K$, which introduces a negative competitive
coupling $\propto - f_1 f_2$.
Notice that for equal dispersal and growth properties, the total
population density $f = f_1 + f_2$ follows a standard
diffusion-advection-growth process,
\begin{eqnarray}
\frac{\partial f}{\partial t}
=
- \nabla \cdot [ \, \mathbf{F} f ]
+  \nabla \cdot (D \nabla f)
+ \alpha f \left( 1 - \frac{f}{K} \right) \, .
\end{eqnarray}

The results of Ref.~\cite{Patriarca2009a} can be summarized as follows.
First, inhomogeneities in the initial distributions are crucial
for the final state, e.g. broader initial distributions represent a
disadvantage for small population growths ($\alpha \to 0$)
while they can become an advantage for large enough values of $\alpha$.
Secondly, also boundary conditions are relevant in the competition process,
e.g. the vicinity of reflecting boundaries definitely favors the survival
of a language for low growth rates.
Finally, geographical barriers such as mountains or rivers can create a
refugium where a linguistic population with lower status can survive
with a stable finite density, even in the presence of an in-flux of
speakers of a higher-status language.

As in the previous section, similar considerations apply also to these
specific models describing human dispersal about possible more general
interpretations as models of dispersal in social space, e.g., through
different social classes.


\section{A Game-Theoretical Model: Bilinguals as a Minority Population}

\subsection{Introductory remarks}
\label{JRintro}

The AS model~\cite{Abrams2003a}, 
where the languages $A$ and $B$ compete for speakers, is
not a good analytical tool to study most of the multilingual societies, as
we know them today. The AS model seems to be more appropriate to describe
the language competition that took place during the historical period of
emergence of nation-states, typically during the 18th and 19th century in
Europe. This is the period in which the nationalist program is about to be
accomplished: the consolidation of a national market, with precise borders,
free of internal restrictions to economic activities, and an increasing
demand for a unique national language that would facilitate those activities
and help develop a national culture and
identity~\cite{Hobsbawm1992a}. 
Some societies, nevertheless, in Europe and elsewhere, 
resisted that shift to language uniqueness. 
Extensions of the AS model have been developed to
understand the existence of those multilingual societies 
(see sections \ref{sec:PL} and \ref{sec:PH}, 
and Refs. \cite{Minett2008a,Mira2005a,Patriarca2004c,Pinasco2006a}). 
The Lotka-Volterra type of models of
language shift \cite{Kandler2008a,Kandler2010a} 
describe fairly well the historical shift to English
of Scottish Gaelic and Welsh, that converted these two vernacular languages
into two minority languages.

Here, we shall consider a society with two languages, $A$ spoken by all its
members, and $B$ spoken by a small proportion $\alpha $. Thus, $\alpha $
denotes the proportion of bilingual speakers and $(1 \!-\! \alpha)$ that of the
monolingual ones. As examples of this situation, we can consider: in Wales,
Welsh and English; in Scotland, Scottish Gaelic and English; in the Basque
Country, Basque and French in the French part and Basque and Spanish in the
Spanish part; in Brittany, Breton and French; Sami and Swedish, Norwegian
and Russian in the Sami society; Frisian, spoken in the province of
Friesland in The Netherlands, competing with Dutch; Maori and English in New
Zealand and Australia; Native American languages (Quechua, Aymar\'{a},
Guarani, among others) and English, Spanish, French, Portuguese and Dutch in
America; languages from the Russian Federation competing with Russian. See
Ref. \cite{Fishman2001a} for more examples.

We could either assume a population of constant size or allow for changes in
the size. In the latter case, any new individual added to the society
(say an immigrant) would, most likely, learn at least $A$.
But in both cases the proportion of individuals who would speak $A$,
in the type of
society we are dealing with, would always be almost, or just, 100\%. Hence,
in those societies, there is no dynamics for language $A$ of any relevance. 
\textit{Only the dynamics of language B matters; and,
therefore, to keep diversity, only the use of B matters.}

We will investigate the language conventions of bilingual speakers by means
of game-theoretic tools. As a methodological procedure, we might think that
our analysis will deal with the stable equilibria obtained by some of the
models used in Refs. \cite{Patriarca2004c,Mira2005a,Pinasco2006a,Minett2008a}, 
where it is formally shown that 
\textit{A} and \textit{B} may coexist. We want to study the language used in
the interactions between bilingual speakers that occur outside the
traditional geographical areas of \textit{B} studied in section \ref{sec:PL}
\cite{Patriarca2004c},
and ask: \textit{will the language conventions developed by the bilingual
speakers use the minority language B and therefore keep up the language
diversity? }

\subsection{The Language Conversation Game (LCG): Iriberri-Uriarte Model}

The model proposed by Iriberri and Uriarte [13] consists of a game played by
two individuals (at least one should be bilingual) to decide the
language that will be used in the conversation that takes place during an
interaction.
The model satisfies the following set of assumptions.

\subsubsection{Assumptions}

\begin{description}

\item[\textbf{Assumption 1 (A.1)}.] 
\textit{Imperfect information}:
Nature or Chance chooses
first the actual realization of the random variable that determines the type
of each speaker (i.e. bilingual or monolingual). But each speaker knows only
his own type. That is, a bilingual speaker does not know, ex-ante, the
bilingual or monolingual type of the agent she will interact with.
We assume, on the other hand, that the probability distribution, $\alpha $
and $(1-\alpha )$, is common knowledge among all the agents in the society

\item[\textbf{Assumption 2 (A.2)}] 
\textit{Linguistic Distance}:
A and B are linguistically
very distant, so that successful communication is only possible when the
interaction takes place in one language.

\item[\textbf{Assumption 3 (A.3)}] 
\textit{Language loyalty}:
Bilingual speakers prefer to use \textit{B}.

\item[\textbf{Assumption 4 (A.4)}]
\textit{Payoffs}:
For a given
proportion $\alpha < 1 - \alpha$, 
we assume the following payoff ordering: $ m > n > c > 0 $.
The maximum payoff $m$ is obtained when bilingual speakers coordinate in
their preferred language B. Bilingual or monolingual players might
coordinate on the majority language $A$; in that case, we will assume both
players get payoffs equal to $n$, because this was a voluntary coordination
or choice. Then $(n-c)$ is the payoff to a bilingual player who, having
chosen $B$, is matched to someone monolingual (or bilingual) who uses
language $A$ and is therefore forced to speak $A$; $c$ denotes the
frustration cost felt by this bilingual speaker.

\item[\textbf{Assumption 5 (A.5)}] 
\textit{Frustration Cost}:
$c<(m-n)\alpha/(1-\alpha )$. 
The bilingual's frustration cost is smaller than the weighted benefit.

\end{description}

\subsubsection{Discussion of the assumptions}

Notice that 
\textbf{A.1} does not allow the existence of a geographical linguistic
partition, as it is assumed in Ref. \cite{Patriarca2004c}, 
since inside the historical areas
where \textit{B} is widely used, there will exist almost perfect information
about the bilingual or monolingual nature of their inhabitants. \textbf{A.1}
tries to capture the use of $B$ outside the strongholds of $B$, in, so to
say, the urban domains. We should take into account that, often, one of the
consequences of a language contact situation is that even the accents, as
signals that would reveal who speaks $B$ and who does not, are erased. If we
eliminate \textbf{A.1 }and assume perfect information and the rest of
assumptions, it is easy to see that bilingual speakers would coordinate in $%
B $. Assumption \textbf{A.2} avoids the linguistic similarities between 
\textit{A} and \textit{B} assumed in Ref. \cite{Mira2005a}. 
If we assume language
similarity, then conversations could take place using both $A$ and $B$,
bilingual speakers would not be forced to change necessarily 
from $B$ to $A$, 
and there would not be any frustration cost. If, instead of \textbf{A.5,}
we assume $c \geq (m-n)\alpha/(1-\alpha )$ while keeping the
other assumptions, then it can be shown that the language used in
equilibrium would be $A$.
See also Ref. \cite{Iriberri2012a}.

\subsubsection{Pure Strategies}

Let us consider a bilingual speaker who must decide, under imperfect
information, which language is going to use in an interaction which is about
to occur. Simplifying things, we could say that the bilingual speaker
expects to be involved in two exclusive events:
\textit{a matching with another bilingual speaker} 
(which will occur with probability $\alpha$) 
and \textit{a matching with a monolingual speaker} 
(which will occur with probability $1-\alpha$). 
We could also simplify the set of
strategies that any bilingual speaker may play to the following two:

\begin{description}

\item[$\mathbf{s}_{\mathbf{1}}$:] 
\textit{Use always B, whether you know
for certain you are speaking to a bilingual individual or not}.

\item[$\mathbf{s}_{\mathbf{2}}$:] 
\textit{Use B only when you know for
certain that you are speaking to a bilingual individual; 
use A, otherwise.}

\end{description}
Note that strategy $\mathbf{s}_{\mathbf{1}}$ reveals the bilingual nature of
the speaker. Strategy $\mathbf{s}_{2}$, on the other hand, hides the
bilingual nature of the speaker. Its purpose is to avoid the frustration
cost $c$ of assumption \textbf{A.4}.

In the present paper, we study the normal form of the LCG. A complete
description of the LCG is given by the extensive form presented in
Ref. \cite{Iriberri2012a}.

The language associated to each pure strategy profile is given by the
following matrix:

\begin{equation*}
\begin{tabular}{ccc}
& $\mathbf{s}_{1}$ & $\mathbf{s}_{2}$ \\ \cline{2-3}
$\mathbf{s}_{1}$ & \multicolumn{1}{|c}{$\mathbf{B}$} & \multicolumn{1}{|c|}{$%
\mathbf{B}$} \\ \cline{2-3}
$\mathbf{s}_{2}$ & \multicolumn{1}{|c}{$\mathbf{B}$} & \multicolumn{1}{|c|}{$%
\mathbf{A}$} \\ \cline{2-3}
\end{tabular}
\end{equation*}

Note that since players have no perfect information about the type of the
opponent, they might use either \textit{B} or \textit{A} in the interaction.
That is, if both bilingual speakers hide their type by choosing 
$\mathbf{s}_{\mathbf{2}}$, then they will use in the interaction their less preferred
language \textit{A}. In the other three cases they will use \textit{B},
because at least one speaker is revealing the bilingual identity.

\subsubsection{Expected Payoffs}

\begin{figure}[t]
\begin{center}
 \includegraphics[width=0.48\textwidth]{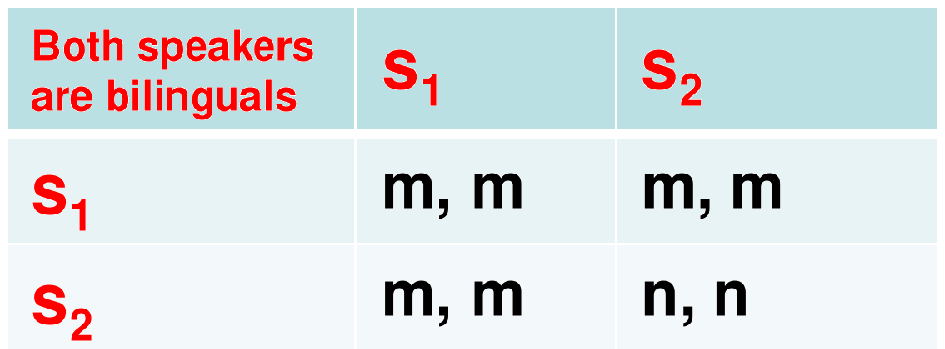}
\end{center}
 \caption {The game to be played in the
 event in which two bilingual speakers are matched. Notice that $s_{1}$ is
 weakly dominant.}
 \label{JRfig1}
\end{figure}

The (mythical) player called Nature or Chance chooses, with
probability $\alpha $, that the bilingual speaker interacts with
another bilingual speaker, to play the game described in
Fig. \ref{JRfig1}, in which, by assumption \textbf{A.4}, 
strategy $\mathbf{s}_{\mathbf{1}}$ is weakly dominant.

\begin{figure}[t]
\begin{center}
 \includegraphics[width=0.48\textwidth]{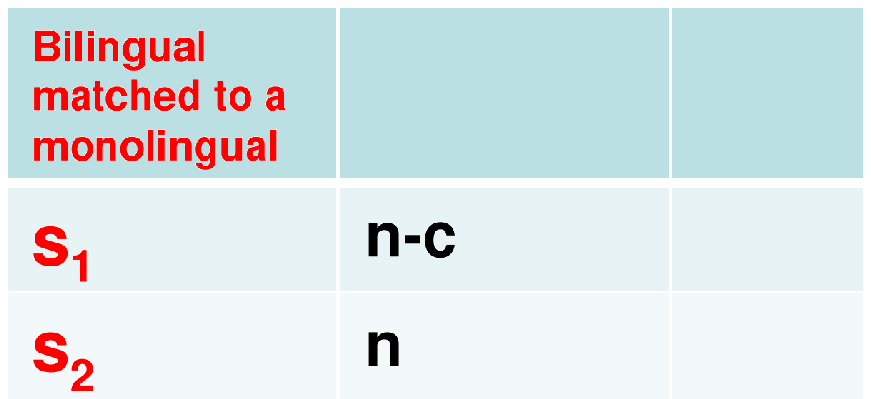}
\end{center}
 \caption {The game to be played in the
event in which a bilingual speaker is matched to a monolingual. Notice
that $s_{2}$ is strictly dominant.}
 \label{JRfig2}
\end{figure}

Nature chooses, with probability $1-\alpha $, that the bilingual speaker
interacts with a monolingual speaker, to play the game described in
Fig. \ref{JRfig2},
in which, by assumption \textbf{A.4}, 
$\mathbf{s}_{\mathbf{2}}$ is strictly dominant.
The monolingual agent does not make choices and gets $n$.

If a bilingual speaker chooses strategy $\mathbf{s}_{\mathbf{1}}$, then, no
matter the choices of the other bilingual player, the expected payoff
is 
$\alpha m+(1-\alpha )(n-c)$; if the choice is $\mathbf{s}_{\mathbf{2}}$,
then, against $\mathbf{s}_{\mathbf{1}}$, 
the expected payoff will be $\alpha m + (1-\alpha) n$ and, 
against $\mathbf{s}_{\mathbf{2}}$, $n$.
The resulting matrix of expected payoffs of the LCG played by two bilingual
speakers  will, therefore, be symmetric:

\begin{equation*}
\begin{tabular}{ccc}
& $\mathbf{s}_{\mathbf{1}}$ & $\mathbf{s}_{\mathbf{2}}$ \\ \cline{2-3}
$\mathbf{s}_{\mathbf{1}}$ & \multicolumn{1}{|c}{$\alpha (m-n)-c(1-\alpha )$, 
$\alpha (m-n)-c(1-\alpha )$} & \multicolumn{1}{|c|}{$\alpha (m-n)-c(1-\alpha
)$, $\alpha (m-n)$} \\ \cline{2-3}
$\mathbf{s}_{\mathbf{2}}$ & \multicolumn{1}{|c}{$\alpha (m-n)$, $\alpha
(m-n)-c(1-\alpha )$} & \multicolumn{1}{|c|}{$0$,$0$} \\ \cline{2-3}
\end{tabular}%
\end{equation*}

\begin{center}
{\small Matrix of Expected Payoffs}
\end{center}

\subsection{Evolutionary Setting}

The LCG will now be viewed as a population game. To this end, let us assume
that the bilingual population consists of a large, but finite number of
individuals, who play a certain pure strategy $s_{\mathbf{i}}$, ($i=1,2$),
in a two-player game. The members of the bilingual population play the LCG
having $S=\{s_{1},s_{2}\}$ as their \textit{common} strategy set. The
interactions are modelled as pairwise random matching between agents of the
bilingual population; that is, no more than two (randomly chosen)
individuals interact at a time. The interactions take place continuously
over time. Let $N$ be the total population of bilingual speakers in the
society, and $x=\frac{\mathbf{N}_{1}}{\mathbf{N}}$ the proportion of
bilingual agents playing the pure strategy $s_{\mathbf{1}}$ at any point 
\textit{t} in time (time dependence is suppressed in the notation). In this
setting, a mixed strategy is interpreted as a population state that
indicates the bilingual population share of agents playing each pure
strategy. On the other hand, the payoffs of the game should not be
interpreted as biological fitness, but as utility. Under assumptions
\textbf{A.1-A.5}, we get the following result.

\bigskip \textbf{Proposition. }
\textit{There exists a mixed strategy Nash
equilibrium in which the bilingual population plays $s_{\mathbf{1}}$
with probability
$x^{\ast }=1-\frac{c(1-\alpha )}{\alpha (m-n)}$.
This equilibrium is evolutionary stable
--- that is $x^{\ast }$
is a language convention built by the bilingual population --- 
and asymptotically stable 
in the associated one-population Replicator Dynamics.}

Proof: Note that the LCG has the strategic structure of a \textit{Hawk-Dove}
Game (with $\mathbf{s}_{\mathbf{1}}$ as \textit{Dove} 
and $\mathbf{s}_{\mathbf{2}}$ as \textit{Hawk}). 
Thus, it has three Bayesian Nash equilibria:
the asymmetric (and unstable) equilibria 
$(\mathbf{s}_{\mathbf{2}},\mathbf{s}_{\mathbf{1}})$ and 
$(\mathbf{s}_{\mathbf{1}},\mathbf{s}_{\mathbf{2}})$, and
the symmetric mixed strategy equilibrium 
$(x^{\ast },1-x^{\ast })=$ $(1-\frac{c(1-\alpha )}{\alpha (m-n)},\frac{c(1-\alpha )}{\alpha (m-n)})$, 
with $x^{\ast }\in (0,1)$. To see that the latter equilibrium is evolutionary
stable, see Ref.~\cite{Weibull1995a}. 
The single population Replicator Dynamics
is as follows: 
\begin{equation*}
\overset{\cdot }{x}\text{ }=[\alpha (m-n)(1-x)-c(1-\alpha )]x(1-x) \, .
\end{equation*}

Notice that in $x^{\ast }$, 
$\alpha (m-n) (1-x) - c (1-\alpha) = 0$ and so 
$\overset{\cdot }{x}=0$. 
We can see that for any 
$ 0 < x < 1 - \frac{c(1-\alpha )}{\alpha (m-n)}$, 
$\overset{\cdot }{x}$ increases toward $x^{\ast }$, and for
any $ 1 > x > 1 - \frac{c(1-\alpha )}{\alpha (m-n)}$, 
$\overset{\cdot }{x}$ decreases toward $x^{\ast }\blacksquare$.

\subsubsection{Interpretation}

Language $B$ is spoken in the asymmetric equilibria 
$(\mathbf{s}_{\mathbf{2}},\mathbf{s}_{\mathbf{1}})$ 
and $(\mathbf{s}_{\mathbf{1}},\mathbf{s}_{\mathbf{2}})$, 
but these equilibria are unstable and, hence, we must rule
them out. Thus, we are left with the evolutionary and asymptotically stable
equilibrium $x^{\ast }$. 
Since $x^{\ast }\in (0,1)$, 
$\mathbf{s}_{\mathbf{1}}$ 
and $\mathbf{s}_{\mathbf{2}}$ are played by non-zero proportions of
bilingual speakers. 
That is, the bilingual population is optimally partitioned in two subpopulations: 
$N_{1}^{\ast }=N[1-\frac{c(1-\alpha )}{\alpha (m-n)}]$ 
and $N_{2}^{\ast }=N\frac{c(1-\alpha )}{\alpha (m-n)}$. 
The former group is composed of agents
who play strategy $s_{\mathbf{1}}$ and the latter of those who play 
$s_{\mathbf{2}}$. Hence, the bilingual agents in $N_{2}^{\ast }$ do
not use $B$ in the interactions among themselves. 
Only when they interact with agents
of $N_{1}^{\ast }$ will they use $B$.

Thus, in equilibrium, the population of bilingual speakers will speak
both $A$ and $B$ in the interactions between themselves; 
the level of use of $B$ depends
on the relative size of $N_{1}^{\ast }$.

\subsection{Concluding remarks}

The mixed equilibrium $x^{\ast }$ is compatible with almost all the possible
levels of use of $B$: from the lowest,
when $x^{\ast }$ is almost $0$
and therefore $N_{2}^{\ast }$ approaches $N$, so that bilingual speakers
will mostly speak $\mathit{A}$ between them;
to the highest,
when $x^{\ast }$ is almost $1$ and so bilingual speakers will be almost
all speaking their preferred language $B$. 
Hence we cannot give a sharp
answer to the question posed in the introductory remarks in
Sec. \ref{JRintro}. 
We can only say that
language diversity is not safe in this equilibrium.

Our prediction is that there is a tendency towards a situation in which $%
N_{2}^{\ast }>N_{1}^{\ast }$, and that the use of \textit{B} will always
face the danger of being reduced to marginal levels outside the traditional
areas. Many factors will intervene, some of them from outside the model.
Among others, imperfect information, the dominance of \textit{A} in formal
and informal usages of the language, and the politeness norms
that would advice the use of (the \textit{Hawk} strategy) 
$\mathbf{s}_{\mathbf{2}}$. 
Bilingual speakers playing $\mathbf{s}_{\mathbf{2}}$ will hurt each
other because they end up speaking the less preferred  language (see the
language matrix above). This might explain the difficulties observed by
Fishman (2001)~\cite{Fishman2001a}. The actual lower bound to the use 
of \textit{B} will be
near to that set by the communities living in the geographical areas where $%
\mathit{B}$ is strong and where interactions occur with almost perfect
information \cite{Patriarca2004c}. 
See Ref. \cite{Iriberri2012a} for a more complete analysis.


\section{Conclusions}

We have revisited several approaches used to study the dynamics of two
competing languages. The main question addressed is whether the
competition leads to the coexistence or on the contrary to the
prevalence of a majority language. The seminal work of the AS model
considers speakers of two languages without the possibility of
bilinguals. In this case, the stationary configurations depend on the
volatility, that is, how easy is to change language use depending on
the local density of speakers. When the volatility is high ($a<1$),
i.e., the probability is larger than the linear case, coexistence is
the stationary solution where the percentage of speakers of each
language depends on the prestige.  When the volatility is low ($a>1$),
the systems converges to the dominance of one of the languages and the
extinction of the other. The presence of bilinguals and network of
interactions (Bilinguals model) change the boundaries separating the
different regimes, but the overall picture remains similar. Physical
and/or political boundaries have been shown to allow for coexistence
as long as the two communities are separated. There have been also
attempts to show the coexistence of monolingual speakers in population
dynamic models and the stability of bilingual communities. Finally, by
means of a game theoretical approach, we have analyzed the case of
language competition when one of languages is known by all the agents
while the other is only spoken by a minority.

Despite the efforts to understand the different mechanisms of language
competition, an overall clear picture on the question of coexistence
(or not) or multilingual communities is still missing. In this respect
empirical works should provide evidence and guidance to improve
current models. The original data in Ref.~\cite{Abrams2003a} on the
evolution of the total number of speakers has triggered the research
line so far. Recent research~\cite{Kandler2010a}
using empirical data on Britain's Celtic languages with good spatial
and temporal resolution should be taken as a motivation for further
studies. We anticipate that future research will address how
spatio-temporal patterns emerge from the competition of local
interaction with global signals (e.g., prestige, language policies).


\begin{acknowledgments}
We acknowledge financial support from the Spanish Ministry of Science
and Innovation MICINN and FEDER through projects ECO2009-11213-ERDF,
FISICOS (FIS2007-60327); MODASS (FIS2011-24785), and SEJ2006-05455;
the Basque Government through project GV-EJ: GIC07/22-IT-223-07; the
Estonian Ministry of Education and Research through Project
No. SF0690030s09 and the Estonian Science Foundation via grant
no. 7466. Jos\'e Ram\'on Uriarte wants to thank the Department of Economics of
Humbolt-Universtät zu Berlin, where this research was completed, for
the facilities offered. We also thank Federico Vazquez for his
contribution to the original work reviewed here.
\end{acknowledgments}

\bibliography{acs-review}
\bibliographystyle{apsrev}

\end{document}